\documentclass[twocolumn,superscriptaddress,showpacs,amsmath,amssymb,aps]{revtex4-1}
\usepackage{hyperref}
\hypersetup{
setpagesize=false,
 bookmarksnumbered=true,%
 bookmarksopen=true,%
 colorlinks=true,%
 linkcolor=blue,
 citecolor=blue,
 urlcolor=blue,
}
\usepackage[dvipdfmx]{graphicx}
\def\average#1{\langle #1 \rangle }

\def\vector#1{\mbox{\boldmath $#1$}}

\def\defeq{\mathrel{\mathop:}=}%
\def\italic#1{{\it #1}}

\def\refeq#1{Eq.\,(\ref{#1})}
\def\reffig#1{Fig.\,\ref{#1}}

\def\itGamma{{\it \Gamma}}%
\def\substitute#1{\left. #1 \right|}

\begin{document}
\title{Localized Mode and Nonergodicity of a Harmonic Oscillator Chain}
\author{Fumihiro Ishikawa}
\affiliation{Department of Physics, The University of Tokyo, Tokyo 113-0033, Japan}
\author{Synge Todo}
\affiliation{Department of Physics, The University of Tokyo, Tokyo 113-0033, Japan}
\affiliation{Institute for Solid State Physics, The University of Tokyo, Kashiwa 277-8581, Japan}
\date{\today}

\begin{abstract}
  We present a simple and microscopic physical model that breaks the ergodicity.
  Our model consists of coupled classical harmonic oscillators, and the motion of the tagged particle obeys the generalized Langevin equation satisfying the second fluctuation dissipation theorem.
  It is found that although the nonergodicity strength, which is expected to detect the ergodicity breaking, for this model vanishes, the velocity auto correlation function of the tagged particle asymptotically oscillates.
  We analyze the model by using the molecular dynamics and the exact diagonalization as well as the rigorous mapping to the generalized Langevin equation.
  Our analysis reveals that the asymptotic oscillation is caused by a localized mode with an isolated frequency from the continuous phonon spectrum.
\end{abstract}

\maketitle 

\section{INTRODUCTION}
The central question of the statistical physics, {\em how a system thermalizes}, is still irresistible.
For the unitary dynamics of the quantum mechanical systems, the eigenstate thermalization hypothesis has been proposed and is being studied actively~\cite{Srednicki1994,Palma-et-al2015,Mori-Shiraishi2017}.
In classical statistical systems, the concept of {\em ergodicity} is also one of the most fundamental problems. Recently, the concept is attracting much attention in the research of the non-Markovian Brownian motion that sometimes shows anomalous diffusion behavior.
In the non-Markovian dynamics, the memory effect of the bath, which is ignored in the Markovian counterpart, is taken into account, and the dynamics is described by the generalized Langevin equation (GLE)~\cite{Mori1965}.
Originally, the GLE was considered too theoretical, but later non-Markovian dynamics following the GLE have been observed in various numerical simulations as well as in experiments~\cite{Kheifets2014,Mo-Simha-Riazen2015,Lesnicki2016,Li2010,Kou2004,Mason1995,Williams-Bryant-Snook2006}. Also, it has been shown that the GLE is written explicitly in some cases~\cite{Yang2003,Clercx-Schram1992}. Even in the water, surprisingly, the Brownian motion exhibits non-Markovian dynamics~\cite{Clercx-Schram1992}. Furthermore, the GLE has been applied to the quantum Brownian motion~\cite{Cohen-Rabani2011,Calderia-Leggett1983} as well as to the molecular dynamics of open systems~\cite{Lorenz-Kantoro2014}.

Here, we should notice that there is some confusion in the recent discussion of the ergodicity. One of the causes is in the definition of the ergodicity. In previous literatures, there are at least two different definitions of the ergodicity. The conventional definition of the ergodicity, that is, {\em the long time average is equal to the ensemble average}, is considered under a particular situation, where the initial configuration of the system is in the thermal equilibrium. For example, in Refs.~\cite{KhinchinThoremText,Kubo1957,Lee2001}, they considered the conventional ergodicity and proposed some criteria for detecting the ergodicity of the system. Considering the normal, i.e., Markovian, Langevin equation, on the other hand, the ergodicity under {\rm arbitrary} initial conditions of the tagged (or focused) particle is often considered. The latter ergodicity is differently defined as follows: {\em the distribution function of the velocity of the tagged particle relaxes to the equilibrium one irrespective of the initial condition}. This definition is expressed as
\begin{align}
  \lim_{t \to \infty} P(v,t | v_0 , t_0 = 0) = P_{\rm eq}(v),
  \label{eqn:2nd_ergodicity}
\end{align}
where $P(v,t | v_0 , t_0)$ is the transition probability of the velocity of the tagged particle, and $P_{\rm eq}(v)$ is the equilibrium distribution. The former ergodicity holds in this case as well. In some recent researches~\cite{Bao2005,Lapas2008}, they considered the latter ergodicity and established some criteria. We consider the ergodicity of the latter definition in the present work.

The GLE is defined as
\begin{align}
m\frac{dv}{dt} = - m\int_{0}^{t} {\it \Gamma}(t-s) v(s) \, ds + R (t) \label{langevin},
\end{align}
where $m$ is the mass of the tagged particle, $\italic{\Gamma}(t)$ is the memory function, and $R(t)$ is the random force. The last two quantities are related with each other through the second fluctuation-dissipation theorem~\cite{Kubo1966,Mori1965}:
\begin{align}
\average{R(t)R(0)} = mk_{\rm B}T\italic{\Gamma}(t),
\end{align}
where $k_{\rm B}$ is the Boltzmann constant and $T$ is the temperature of the bath.
The GLE~(\ref{langevin}) is derived formally by the projection method~\cite{Zwanzig1960,Nakajima1958,Mori1965,Mori-Fujisaka1973,Zwanzig1961} or the continued fraction method~\cite{Lee1982,Lee1982-2}.
For the system described by the GLE, it is known that the transition probability of the velocity behaves as~\cite{Adelman1976}
\begin{align}
\begin{split}
& P(v,t | v_0,t_0 = 0) \propto \exp \biggl( - \frac{m}{2k_{\rm B} T}\frac{ \bigl[ v - v_0 \, a(t)\bigr]^2}{1 - a^2(t)} \biggr),
\end{split}
\end{align}
where $a(t)$ is the velocity auto correlation function (VACF) of the tagged particle:
\begin{align}
a(t) \defeq \frac{ \average{v(t)v(0)}}{ \average{v^2(0)}}.
\end{align}
with $v(0)=v_0$.
Here, $\average{\cdot}$ means the average over the initial velocity, $v(0)$, and the random force, $R(t)$. According to the projection method, the average over $R(t)$ is equivalent to and thus can be replaced by the average over the initial configuration of the bath degrees of freedom~\cite{Mori1965,Zwanzig1961,Nakajima1958}, where the initial distribution of the bath is assumed to be in the thermal equilibrium. In the discussion of the ergodicity of the latter definition [Eq.~(\ref{eqn:2nd_ergodicity})], the initial distribution of $v(0)$ is arbitrary. The VACF is also the Green function that satisfies
\begin{align}
\frac{da(t)}{dt} = - \int_0^{t} \itGamma (t-s) a(s) \, ds. \label{Green_eq}
\end{align}
According to the definition of the ergodicity we consider, if the tagged particle is ergodic, then
\begin{align}
  \lim_{t \to \infty} a(t) = 0,
\end{align}
irrespective of the initial condition, $v(0)$.

In the meantime, a quantity that could detect the nonergodicity of the GLE has been proposed by several authors~\cite{Costa-et-al2003,Bao2005,Bao2006,Lapas-et-al2007}. The quantity, which is denoted by $b$, is called the ``nonergodicity strength,'' and defined by
\begin{align}
  b \defeq  \Big[ 1 + \displaystyle \lim_{z \to 0} \frac{\overline{{\it \Gamma}}(z)}{z}\Big]^{-1}.
  \label{non-ergodic-strength}
\end{align}
Here, $\overline{\italic{\Gamma}}(z)$ is the Laplace transform of the memory function $\italic{\Gamma}(t)$. According to Refs.~\cite{Bao2005,Bao2006,Lapas-et-al2007}, $b>0$ ($b=0$) would indicate that the system is nonergodic (ergodic) in the present definition.

The nonergodicity strength has been introduced by considering the convergence of the VACF of the tagged particle
\begin{align}
\begin{split}
  \lim_{t \to \infty} a(t) = \lim_{z \to 0} z \overline{a}(z) = \Big[ 1 + \displaystyle \lim_{z \to 0} \frac{\overline{{\it \Gamma}}(z)}{z} \Big]^{-1},
  \label{eqn:10}
\end{split}
\end{align}
where the first equality is derived from the final value theorem, and
the second one is by using the Laplace transform of $a(t)$ obtained from \refeq{Green_eq}:
\begin{align}
\overline{a}(z) = \frac{1}{z+\overline{\itGamma}(z)}. \label{solution_vacf}
\end{align}
Thus, the vanishing nonergodicity strength would mean the convergence of the VACF, in other words, the nonergodicity is attributed to the anomalous diffusion.

Bao {\it et al.} showed that the nonergodicity strength~(\ref{non-ergodic-strength}) can distinguish the ergodicity breaking in some examples~\cite{Bao2005,Bao2006}. It was pointed out more recently, however, that a particular class of the GLE has a nonergodic solution, even though the integral of the memory function is nonzero~\cite{Plyukhin2011}. In this case, the VACF of the tagged particle oscillates asymptotically.
In Ref.~\cite{Plyukhin2011}, the memory functions that yield the nonergodic behavior have been discussed along with a set of conditions for the GLE to be {\em physical}. However, physical models that have such a memory function have not been known so far.

\begin{figure}[tbp]
  \resizebox{0.4\textwidth}{!}{\includegraphics{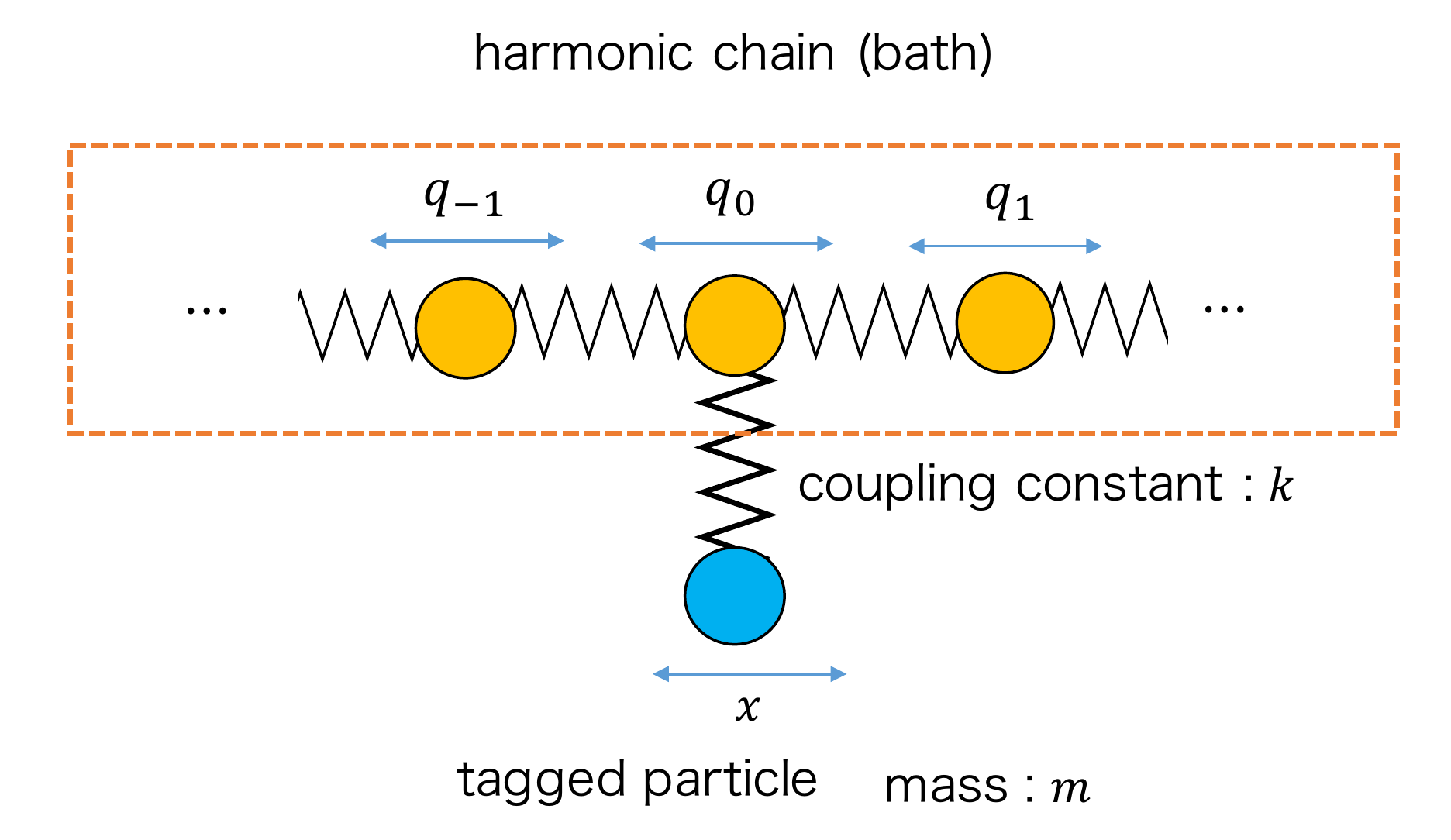}}
  \caption{Schematic picture of our model.
      The harmonic oscillator chain in the orange dashed box is regarded as the {\em bath}, and the tagged particle is connected to the center of harmonic chain with the harmonic interaction.  
      The motion of all the particles is restricted in one dimensional, and they are connected by ideal springs. See Eq~(\ref{eqn:hamiltonian}) for the detailed definition of the model.}
    \label{figure:addRubin}
\end{figure}

In this paper, we present a microscopic physical model that exhibits a nonergodic behavior, even though it has vanishing nonergodic strength, $b=0$. Our model consists of harmonic oscillators. 
The total Hamiltonian of our model is given by
\begin{align}
  \begin{split}
  \mathcal{H} & = \sum_{n = -\infty}^{\infty} \biggl[ \frac{1}{2} \dot{q}^2_n + \frac{1}{2} (q_n - q_{n+1})^2 \biggr]
  + \frac{m}{2}\dot{x}^2 + \frac{k}{2} (x - q_0)^2,
  \end{split}
  \label{eqn:hamiltonian}
\end{align}
where $x$ is the coordinate of the tagged particle, the first term represents a harmonic chain of infinite particles, the second one is the  kinetic energy of the tagged particle, and the third one denotes the interaction between the chain and the tagged particle.
The mass of the particles and the spring constant within the chain are set as unity without loss of generality, while the mass of the tagged particle and the spring constant between the chain and the tagged particle are $m$ and $k$, respectively.
The schematic image of the present model is given in \reffig{figure:addRubin}.
The harmonic chain with infinite length plays a role of the ``thermal bath'' for the tagged particle.
The harmonic oscillator systems are widely used in analysis of the GLE~\cite{Isao2000,Rubin1963,Ford-Kac-Mazur1964,Adelman-Doll1976, Bao2017, Zwanzig1973}.
The memory function and the VACF of our model can be obtained analytically in the thermodynamic limit.
We demonstrate that the VACF of the tagged particle oscillates asymptotically.
We also investigate the model numerically by the molecular dynamics and the exact diagonalization, and elucidate that the ergodicity breaking is caused by a localized mode with an isolated frequency from the continuous phonon spectrum.

This paper is organized as follows: In Sec.~\ref{sect:analytical}, we derive the analytical solution in the thermodynamic limit. In  Sec.~\ref{sect:numerical}, we introduce our numerical method, the molecular dynamics and the exact diagonalization, and present the numerical results for the present model. Finally, in Sec.~\ref{sect:summary}, we summarize our results and discussion. In Appendix, we present a method to make sampling from the Boltzmann distribution for a harmonic oscillator chain.

\section{ANALYTICAL SOLUTION} \label{sect:analytical}
\subsection{Laplace transform of generalized Langevin equation}
First of all, we consider the GLE of the present model, where the memory function and the VACF can be obtained analytically from the microscopic Hamiltonian~(\ref{eqn:hamiltonian}). In this paper, we derive the GLE by the method used in Ref.~\cite{Adelman-Doll1976}. 

The Laplace transform of a function $f(t)$ is defined as
\begin{align}
  \overline{f}(z) \defeq \int_{0}^{\infty} f(t) e^{-zt} \,dt.
  \label{eqn:laplace}
\end{align}
As for the Laplace transform of derivatives, it is well known that the following relations hold:
\begin{align}
\overline{\frac{df}{dt}} & = z \overline{f}(z) - f (0) \label{eqn:laplace-derivative} \\
\overline{\frac{d^2f}{dt^2}}  & = z^2 \overline{f}(z) -  \dot{f} (0) - z f (0). \label{eqn:laplace-derivative2}
\end{align}
By taking the Laplace transform, the left-hand side of Eq.~(\ref{langevin}) becomes
\begin{align}
  m\overline{\frac{dv}{dt}} = m\overline{\frac{d^2x}{dt^2}} = m(z^2 \overline{x}(z) - \dot{x} (0) - z x(0)).
\end{align}
On the other hand, since the first term in the right-hand side of Eq.~(\ref{langevin}) is a convolution of $\itGamma (t)$ and $v(t)$, the result of the Laplace transform of the right-hand side becomes
\begin{align}
  \begin{split}
    & -m \overline{\itGamma}(z) \overline{v}(z) + \overline{R}(z) \\
    & \qquad = -mz \overline{\itGamma}(z) \overline{x}(z) + m\overline{\itGamma}(z) x(0) + \overline{R}(z).
  \end{split}
   \label{eqn:gle-right}
\end{align}

\subsection{Memory function}
The analytic solution for the memory function is derived in the form of Laplace transform.
First of all, the equations of motion are derived from the Hamiltonian~(\ref{eqn:hamiltonian}) as
\begin{align}
m\frac{d^2 x}{dt^2}  & = - k(x - q_0) \label{eq:motion1} \\
\frac{d^2 q_0}{dt^2}  & = - k (q_0 - x) -  (q_0 - q_1) - (q_0 - q_{-1}) \label{eq:motion2} \\
\frac{d^2 q_n}{dt^2} & = -  (q_n - q_{n-1}) - (q_n - q_{n+1})   \qquad \text{for $n \neq 0$}. \label{eq:motion3} 
\end{align}
We apply the Laplace transform to these equations.
By taking the Laplace transform of the both sides of Eq.~(\ref{eq:motion3}), and using the relation~(\ref{eqn:laplace-derivative}), we obtain
\begin{align}
  \begin{split}
(z^2 + 2 )\overline{q}_n(z)   & =  \overline{q}_{n-1}(z) + \overline{q}_{n+1}(z) +\dot{q}_n(0)  + z q_n(0) .
 \label{eqn:s10}
   \end{split}
\end{align}
Here, we assume the following relation between $\overline{q}_n(z)$ and $\overline{q}_{n+1}(z)$:
\begin{align}
  \overline{q}_{n+1}(z) = A(z) \overline{q}_n(z) + B_{n+1}(z)
  \label{analytical:relation_1}
\end{align}
for $n \ge 0$. Note that we assume that $A(z)$ does not depend on $n$. If this relations is satisfied, we can rewrite Eq.~(\ref{eqn:s10}) as
\begin{align}
\overline{q}_n(z) & = \frac{\overline{q}_{n-1}(z)}{z^2 + 2 - A(z)} + \frac{\dot{q}_n(0)  + z q_n(0) + B_{n+1}(z)}{z^2 + 2 - A(z)}.
\label{otherlapcorr}
\end{align}
By comparing with Eq.~(\ref{analytical:relation_1}) for $\overline{q}_{n-1}(z)$ and $\overline{q}_n(z)$, we have
\begin{align}
  A(z) &= \frac{1}{z^2 + 2 - A(z)} \label{eqn:S13} \\
  B_n(z) &= \frac{\dot{q}_n(0)  + z q_n(0) + B_{n+1}(z)}{z^2 + 2 - A(z)} \qquad \text{for $n>0$}.
  \label{eqn:bn_positive}
\end{align}
By solving Eq.~(\ref{eqn:S13}), we obtain
\begin{align}
  A(z) = \frac{z ^2+ 2 - z \sqrt{z^2 + 4}}{2},
\end{align}
where we choose one of the two solutions so that the Laplace transform of the memory function becomes positive. Otherwise, it gives an unphysical result. For $n \le 0$, we can consider the similar relation:
\begin{align}
\overline{q}_{n-1}(z) = A(z) \overline{q}_n(z) + B_{n-1}(z) \label{analytical:relation_m}
\end{align}
with the same $A(z)$ and
\begin{align}
  B_n(z) &= \frac{\dot{q}_n(0)  + z q_n(0) + B_{n-1}(z)}{z^2 + 2 - A(z)} \qquad \text{for $n<0$}.
  \label{eqn:bn_negative}
\end{align}

Next, let us consider the Laplace transform of Eq.~(\ref{eq:motion2}):
\begin{align}
  \begin{split}
    & z^2 \overline{q}_{0}(z) - \dot{q}_{0}(0) - z q_{0}(0) = - k \bigl(\overline{q}_{0}(z) - \overline{x}(z) \bigr) \\ & \qquad 
    -  \bigl(\overline{q}_{0}(z) - \overline{q}_{-1}(z) \bigr) - \bigl( \overline{q}_{0}(z) - \overline{q}_{1}(z) \bigr).
  \end{split}
\end{align}
By inserting Eqs.~(\ref{analytical:relation_1}) and (\ref{analytical:relation_m}) for $n=0$, we have
\begin{align}
  \overline{q}_{0}(z) & = \frac{k}{k + z \sqrt{z^2 + 4}} \overline{x}(z) + B_{0}(z),
\end{align}
where $B_0(z)$ is defined as
\begin{align}
  B_{0}(z) \defeq \frac{\dot{q}_{0}(0) + z q_{0}(0) + B_{1}(z) + B_{-1}(z)}{k + z \sqrt{z^2+4}}.
\end{align}
Note that according to the recursion relations, Eqs.~(\ref{eqn:bn_positive}) and (\ref{eqn:bn_negative}), $B_0(z)$ is a linear combination of the initial condition $\{q_n(0),\dot{q}_n(0)\}_{n=\infty}^\infty$, but it depends on neither $x(0)$ nor $\dot{x}_n(0)$.

Then, we can write down the Laplace transform of Eq.~(\ref{eq:motion1}) as
\begin{align}
 \begin{split}
& m(z^2 \overline{x}(z) - \dot{x} (0) - z x(0)) = - k \bigl(\overline{x}(z) - \overline{q}_0(z) \bigr) \\
 & \qquad = - \frac{kz\sqrt{z^2+4}}{k + z \sqrt{z^2 + 4}} \overline{x}(z) + k B_0(z). \label{preGLE}
  \end{split}
\end{align}
By comparing with Eq.~(\ref{eqn:gle-right}), we obtain
\begin{align}
  \overline{\itGamma}(z) & = \frac{k\sqrt{z^2+4}}{m(k  + z \sqrt{z^2 + 4})} \label{memoryLap}
\end{align}
and
\begin{align}
  \overline{R}(z) & = -m\overline{\itGamma}(z)x(0) + k B_0(z) .
  \label{randomLap}
\end{align}
From Eq.~(\ref{memoryLap}), we conclude
\begin{align}
\lim_{z \to 0} \overline{{\it \Gamma}}(z) = \frac{2}{m},
\end{align}
i.e., $b = 0$.
Thus, the nonergodic strength indicates that the motion of the tagged particle is {\em ergodic}.

\subsection{Second fluctuation dissipation theorem}
Next, we confirm the second fluctuation dissipation theorem. The second fluctuation dissipation theorem for the GLE is written as
\begin{align}
  \average{R(t)R(0)} = m k_B T \itGamma(t),
\end{align}
and the Laplace transform of the relation yields
\begin{align}
  \average{\overline{R}(z)R(0)} = mk_B T \overline{\itGamma}(z).
  \label{eqn:fdt-z}
 \end{align}
By using \refeq{randomLap}, the left-hand side of \refeq{eqn:fdt-z} is evaluated as
\begin{align}
  \average{\overline{R}(z)R(0)} &  = -m\overline{\itGamma}(z) \average{x(0)R(0)}
  + k \average{B_0(z) R(0)}. \label{eqn:expand1}
\end{align}
From the GLE~(\ref{langevin}), $R(0)$ should satisfy
\begin{align}
  m \substitute{\frac{d^2 x}{dt^2}}_{t =0} & = R(0),
\end{align}
while from the equation of motion we have
\begin{align}
  m \substitute{\frac{d^2 x}{dt^2}}_{t =0}  & = -k x(0) + k q_0(0).
  \label{eqn:S27}
\end{align}
These relations yield
\begin{align}
  R(0) & = -k x(0) + k q_0(0).
\end{align}

We assume the initial condition is in the thermal equilibrium. Since the Hamiltonian is translational invariant, one of the coordinates can be fixed to a constant. Here, we set $q_0(0)=0$ without loss of generality. Then, the Boltzmann distribution function is factorized as
\begin{align}
  \begin{split}
  P(x(0),\dot{x}(0),\{q_n(0),\dot{q}_n(0)\})  = \mathcal{N}(x(0);k_{\rm B}T/k) \delta(q_0(0)) \\
 \times  P_q(\{q_n(0)\}_{n \ne 0}) \mathcal{N}(\dot{x}(0); k_{\rm B}T/m) \prod_n \mathcal{N}(\dot{q}_n(0); k_{\rm B}T),
  \label{eqn:boltzmann}
  \end{split}
\end{align}
where $k_{\rm B}$ is the Boltzmann constant, $T$ is the temperature, $\mathcal{N}(\,\cdot\,;\sigma^2)$ is the Gaussian distribution with mean zero and variance $\sigma^2$, and
\begin{align}
  P_q(\{q_n(0)\}_{n \ne 0}) \propto \exp \bigl [ - \frac{1}{2k_BT} \sum_n (q_n(0) - q_{n+1}(0))^2 \bigr ]
  \label{eqn:boltzmann2}
\end{align}
with $q_0(0) = 0$. Since $x(0)$ decouples with other degrees of freedom and $q_0(0)=0$,
\begin{align}
  \average{\dot{x}(0)R(0)} = \average{q_n(0)R(0)} = \average{\dot{q}_n(0)R(0)} = 0
\end{align}
hold for all $n$, and the second term in Eq.~(\ref{eqn:expand1}) vanishes. Finally, from
\begin{align}
  \average{x(0)R(0)} = -k \average{x^2(0)} = -k_{\rm B}T
\end{align}
we obtain the second fluctuation dissipation relation, Eq.~(\ref{eqn:fdt-z}).

\subsection{Velocity auto correlation function}
Finally, we derive the velocity auto correlation function (VACF) of the tagged particle. From the GLE~(\ref{langevin}), the velocity auto correlation function,
\begin{align}
  a(t) \defeq \frac{\average{v(t)v(0)}}{\average{v^2(0)}},
\end{align}
satisfies
\begin{align}
\frac{da(t)}{dt} = - \int_{0}^{t} \itGamma (t-s) a(s) \, ds.
\end{align}
We apply the Laplace transform to the equation, and obtain
\begin{align}
z\overline{a}(z) - a(0) & = - \overline{\itGamma}(z)\overline{a}(z).
\end{align}
By definition, $a(0) = 1$, then we have
\begin{align}
  a(z) & = \frac{1}{z+\overline{\itGamma}(z)}. \label {analytcal_vacf}
\end{align}
We substitute the Laplace transform of the memory function [Eq.~(\ref{memoryLap})] to Eq.~(\ref{analytcal_vacf}), and then we finally arrive at the explicit form of the VACF:
\begin{align}
\overline{a}(z) & = \frac{m \bigl( k + z \sqrt{z^2 + 4} \bigr) }{ (m z^2 + k) \bigl( \sqrt{z^2 + 4} \bigr) + m k z }.
\end{align}
By the relation between the Fourier and the Laplace transformations, the Fourier transform of the VACF, $\widetilde{a}(\omega)$, is obtained as
\begin{align}
  \begin{split}
    \widetilde{a}(\omega) & = \mbox{Re} \; \overline{a}(-i \omega) \\
    & = \mbox{Re} \Big[ \frac{(m k^2)\sqrt{4 - \omega ^2} }{ (k - m \omega ^2)^2 (4 - \omega ^2 )+ (m k )^2 \omega ^2}  \\
      & \qquad  - i \omega  \frac{ m(k - m \omega^2)(4 - \omega ^2) - m^2 k^2 }{ (k - m \omega^2)^2 (4 -\omega^2 ) + (m k )^2 \omega^2 } \Big].
  \end{split}
   \label{equation:fouier_of_VACF_1}
\end{align}
The first term in the right-hand side of \refeq{equation:fouier_of_VACF_1} is real for $|\omega| \leq 2$.
It should be noted that not only the first term, which gives a continuous spectrum, contributes to $\widetilde{a}$, but also the second term may have a nonvanishing contribution when the denominator becomes zero.
Indeed the denominator in the right-hand side of \refeq{equation:fouier_of_VACF_1} can be zero for any value of $k > 0$, which causes a Dirac delta function in the spectrum through the Cauchy principal value.
Eventually, the spectral density $\widetilde{a}(\omega)$ is given by
\begin{align}
  \begin{split}
    \widetilde{a}(\omega) &= \frac{(m k ^2)\sqrt{4 - \omega ^2} }{(k - m \omega ^2)^2 (4 - \omega ^2 )  + (m k )^2 \omega ^2} \Theta (2-|\omega|) \\
& \qquad + \alpha \, \delta ( \omega - \omega_{\rm iso}),
\end{split}
  \label{equation:fourier_trans_VACF}
\end{align}
where $\Theta(x)$ is the Heaviside step function, $\omega_{\rm iso}$ is the solution of
\begin{align}
  & (k - m \omega_{{\rm iso}} ^2)^2 (4 - \omega_{{\rm iso}} ^2 )  + (m k )^2 \omega_{{\rm iso}} ^2 = 0, \label{equation:condition_of_existence_isolated_mode}
\end{align}
and $\alpha$ is the residue of the pole at $\omega = \omega_{\rm iso}$:
\begin{align}
  \alpha = \lim_{\omega \to \omega_{{\rm iso}}} 2\pi i(\omega - \omega_{{\rm iso}} ) \overline{a}(-i \omega).
\end{align}
It is straightforward to show that $\overline{a}(-i \omega)$ has a simple pole at $\omega = \omega_{{\rm iso}} > 2$; First, the left-hand side of Eq.~(\ref{equation:condition_of_existence_isolated_mode}) takes $4m^2k^2$ $(>0)$ at $\omega^2_{{\rm iso}} = 4$ and $k(1 \pm m)/m$, and becomes negative for $\omega^2_{{\rm iso}} \gg 4$. Therefore, $k(1 - m)/m > 4$ is a necessary condition for Eq.~(\ref{equation:condition_of_existence_isolated_mode}) to have multiple solutions for $\omega^2_{{\rm iso}} > 4$. However, if $k(1 - m)/m > 4$ holds, the discriminant of Eq.~(\ref{equation:condition_of_existence_isolated_mode}) as a cubic equation of $\omega_{{\rm iso}}^2$ is always negative. Thus, Eq.~(\ref{equation:condition_of_existence_isolated_mode}) has just one simple zero at $\omega = \omega_{{\rm iso}} > 2$.

Hereafter, the parameters are set as $m = k = 1$. The position of the pole is evaluated as
\begin{align}
  \omega_{\rm iso} = 2.09355577\cdots
  \label{eqn:omega_iso}
\end{align}
which is isolated from the continuous spectrum given by the first term in \refeq{equation:fourier_trans_VACF}.
Thus, the VACF of the tagged particle has a contribution of the delta function with an isolated frequency in addition to the continuous spectrum function, which means that the VACF does not decay even in the long time limit, and oscillates asymptotically. We conclude that the present model breaks the ergodicity, even though the nonergodicity strength vanishes.

\section{NUMERICAL SIMULATION} \label{sect:numerical}
\subsection{Molecular dynamics}
The inverse Fourier transformation of \refeq{equation:fourier_trans_VACF} is not easy numerically nor analytically.
In order to observe the behavior of the VACF as a function of time in more detail and confirm the validity of our analytic solution, we analyze the present model with a finite chain length, $N$, by using the molecular dynamics simulation.
We consider the periodic boundary conditions, $q_{n+N} = q_{n}$ is imposed.
The total number of the time step is $10^5$, the step size is $10^{-2}$, and thus the total simulation time is $10^3$.  For the numerical integration scheme, we adopt the velocity Verlet method. The initial configuration is sampled from the Boltzmann distribution at temperature $k_{\rm B}T=1$ by using the total shift sampling method (see Appendix~\ref{sect:app}).
We take the average of the time series data over $10^6$ initial configurations.

\begin{figure}[tbp]
  \includegraphics[width=9.5cm]{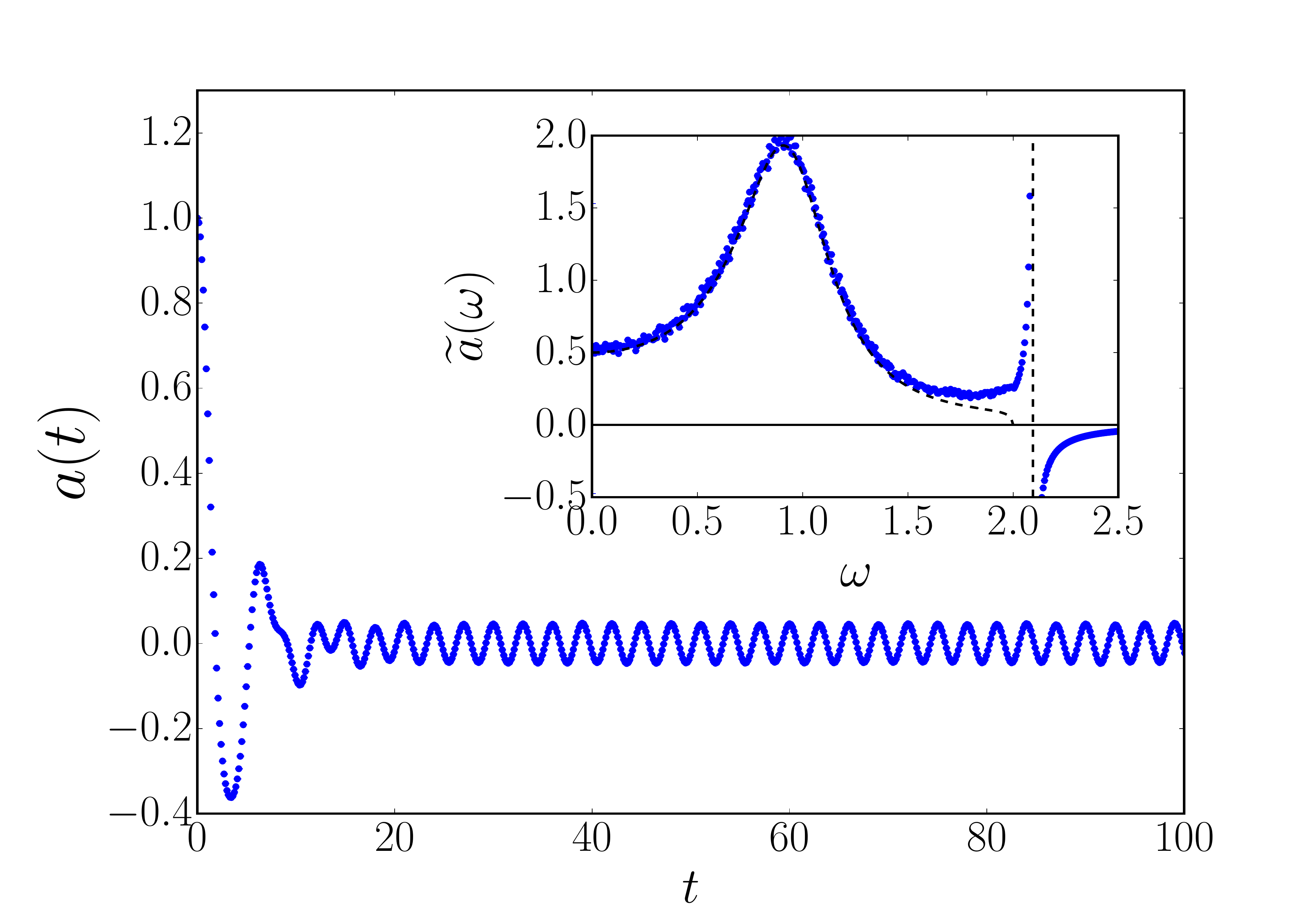}
  \caption{VACF and its spectrum (inset) obtained by the molecular dynamics simulation $N+1=10^4$. The spectrum is obtained by the discrete Fourier transformation. In the inset, the divergence around $\omega \simeq \omega_{\rm iso}$ (indicated by the vertical dashed line) manifests the existence of a delta function. The analytic solution for $\widetilde{a}(\omega)$ [\refeq{equation:fouier_of_VACF_1}] is plotted by the dashed curve in the inset.}
  \label{figure:VACF_and_Spectrum}
\end{figure}

The result of simulation is given in \reffig{figure:VACF_and_Spectrum}. It exhibits a clear oscillation for $t > 20$ in the VACF. The inset in \reffig{figure:VACF_and_Spectrum} shows the spectrum of the VACF, i.e., the discrete Fourier transform of $a(t)$. Near $\omega = 2$, we observe a singular behavior, which manifests a delta peak in the long time limit. The frequency is close to $\omega_{\rm iso}$ obtained from the exact solution [\refeq{eqn:omega_iso}].
Thus, we numerically confirm that the VACF oscillates asymptotically, which is consistent with the analytic solution.

\subsection{Exact diagonalization}
\begin{figure}[tbp]
  \includegraphics[width=8.5cm]{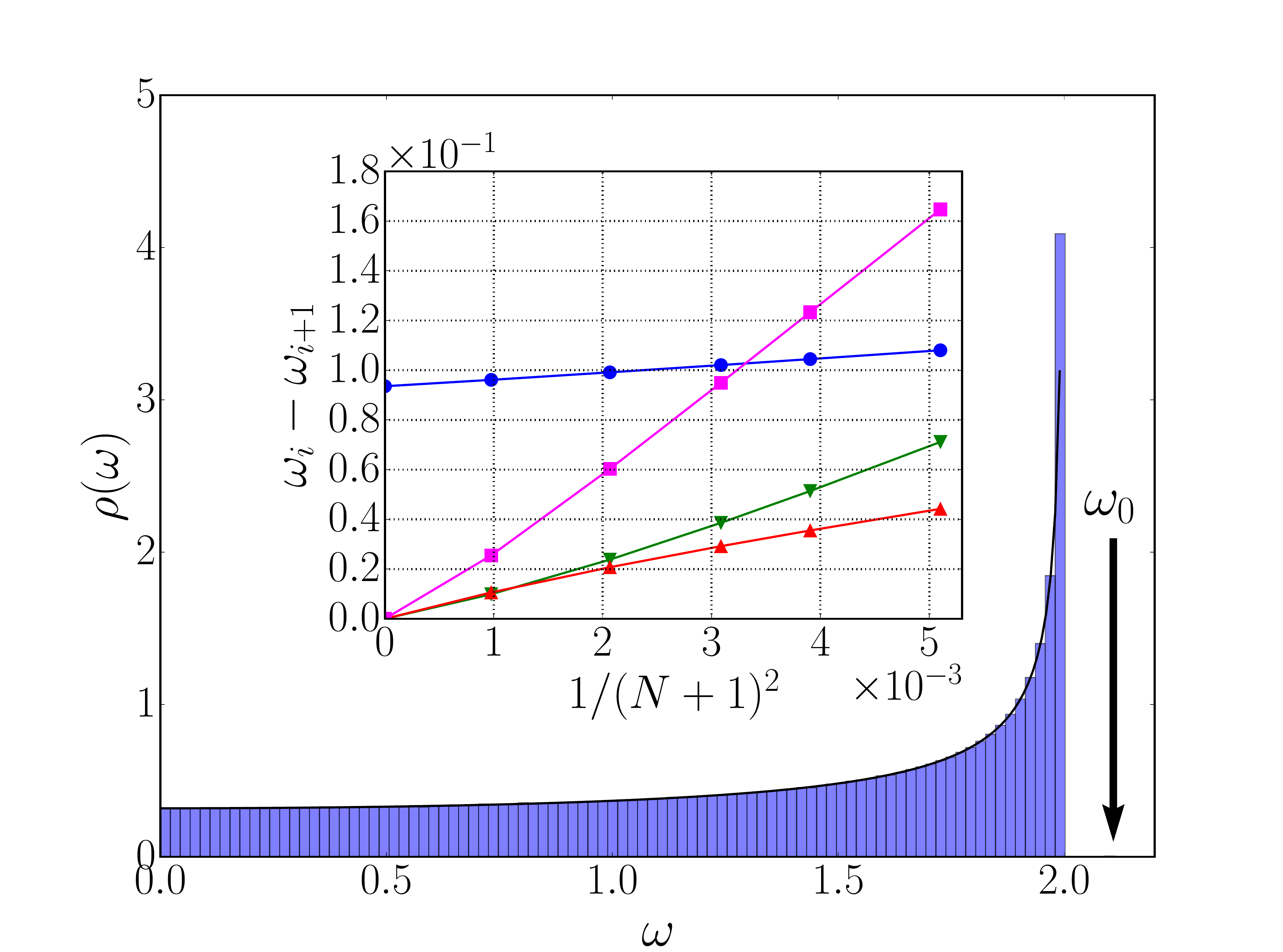}
  \caption{Density of states of the normal modes obtained by the exact diagonalization $N+1=10^4$ (filled boxes). The phonon mode of a simple harmonic chain is represented by the solid line. The arrow (labeled as `$\omega_0$') represents the position of the isolated normal mode with the largest frequency, $\omega_0$. The inset shows the size dependence of $\omega_{i} - \omega_{i+1}$ for $i=0$ (blue circles), $1$ (purple squares), $2$ (green downward triangles), $3$ (red upward triangles). The gap between $\omega_0$ and $\omega_{1}$ in the $N=\infty$ limit is estimated as $0.09355570(4)$ by the least squares method.}
  \label{figure:Spectrum_Number}
\end{figure}

In order to reveal the origin of the oscillation, we further analyze the system by using the exact diagonalization. The total Hamiltonian with $m=k=1$ can be written in the following quadratic form:
\begin{align}
\mathcal{H} = \frac{1}{2} \dot{\vector{q}}^t \dot{\vector{q}} + \frac{1}{2} \vector{q}^t \vector{L} \vector{q},
\end{align}
where $\vector{q} = \{q_0, \cdots , q_{N-1}, x \}$ and $\vector{L}$ is the $(N+1) \times (N+1)$ Laplacian matrix. The square root of the eigenvalue of the Laplacian matrix gives the the frequency of each normal mode, denoted as $\omega_i$ ($i=0,1,\cdots,N$). The eigenvectors represent the amplitude of the eigenmodes, denoted as $\{A_i(n)\}_{i=0}^N$ where $n$ is the sight number and $i$ corresponds to the $i$th eigenvalue. We assume the descending order for the frequencies: $\omega_0 \ge \omega_1 \ge \omega_2 \ge \cdots$.
In \reffig{figure:Spectrum_Number}, we plot the density of state obtained:
\begin{align}
  \rho(\omega) \simeq \frac{1}{N+1} \sum_{i=0}^N \delta(\omega - \omega_i)
\end{align}
by the exact diagonalization for $N+1=10^4$. The continuous spectrum observed for $0 < \omega < 2$ corresponds to the phonon mode of a simple harmonic chain. In addition, at $\omega \simeq 2.09$, an isolated normal mode exists. As shown in the inset of \reffig{figure:Spectrum_Number}, the gap between the largest and the next frequencies, $\omega_0-\omega_1$ converges to a finite value in the $N = \infty$ limit, while the gap between $\omega_i$ and $\omega_{i+1}$ vanishes for $i \ge 1$. By the least squares fitting by using the results whose sizes are from $N+1=10^3$ to $10^4$, we obtain $\omega_0-\omega_1=0.09355570(4)$ for $N \rightarrow \infty$, which coincides with the gap between the isolated mode and the continuum, $\omega_{\rm iso} - 2$, obtained from the analytic solution [Eq.~(\ref{eqn:omega_iso})].

\begin{figure}[tbp]
  \includegraphics[width=8.8cm]{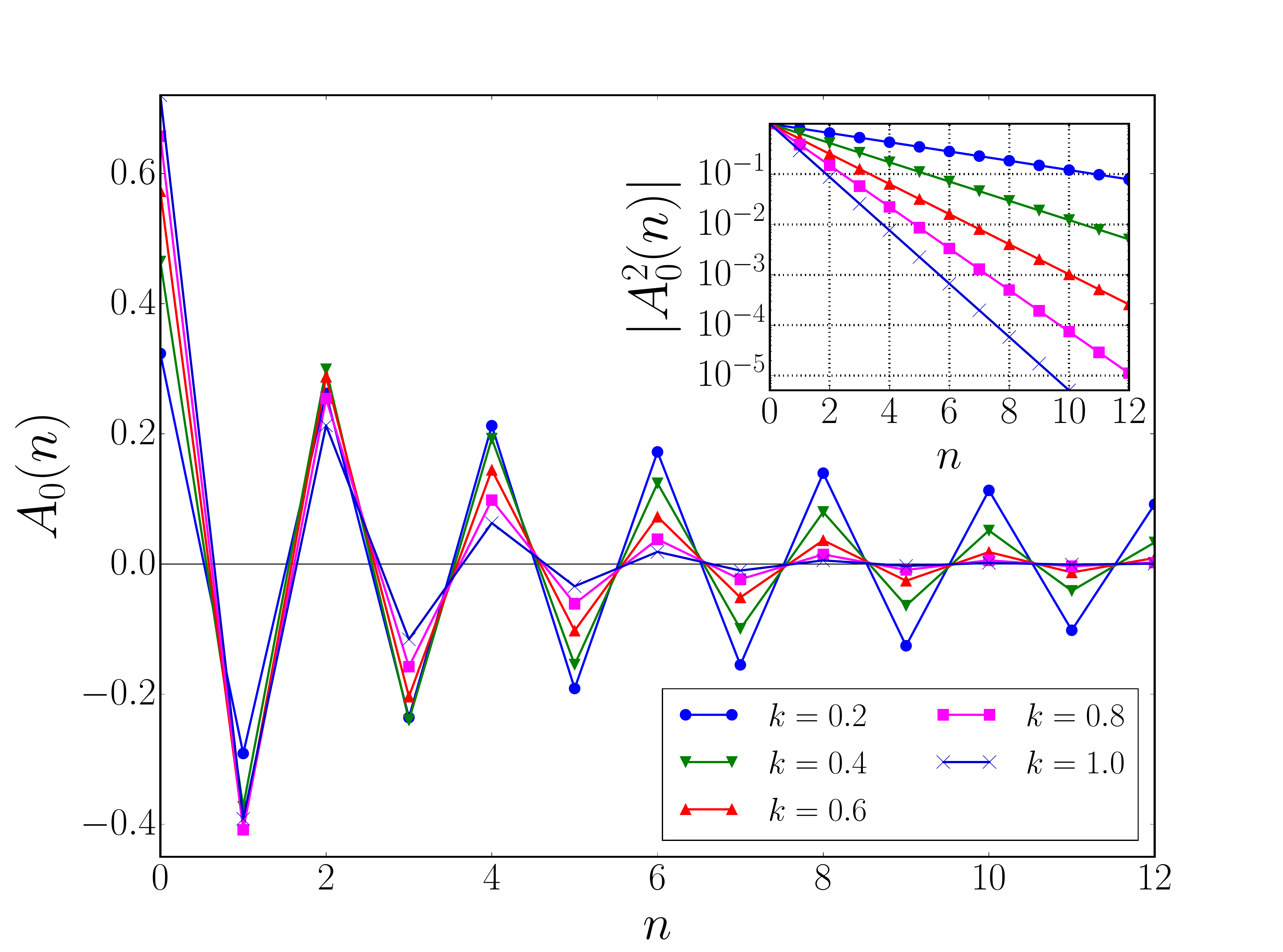}
  \caption{Amplitude of the isolated normal mode, $A_0(n)$, near $n=0$ for $N+1=10^4$, $m=1.0$ and $k=0.2, 0.4, 0.8, 1.0$.
    The horizontal axis, $n$, denotes the site index of the bath particles.  The inset shows a semi-log plot of the amplitude squared.
  }
  \label{figure:Localized_mode}
\end{figure}

In order to assess the spatial structure of the isolated normal mode, next, we plot the amplitude of the normal mode, $A_0(n)$, near $n=0$.
As clearly seen in \reffig{figure:Localized_mode}, the isolated normal mode is localized around $n=0$, at which the oscillator couples to the tagged particle. The inset of \reffig{figure:Localized_mode} shows that the amplitude of the isolated normal mode decays exponentially. From these observations, we conclude that the asymptotic oscillation in the VACF is caused by the isolated and localized normal mode around $n=0$.

\section{SUMMARY AND DISCUSSION} \label{sect:summary}
In the present paper, we proposed a simple physical model of harmonic oscillators [\refeq{eqn:hamiltonian}], in which the dynamics of the tagged particle is described by the GLE and the memory function satisfies the second fluctuation dissipation theorem.
By deriving the analytic solution for the memory function and the VACF, and also by using the numerical simulations, we revealed two notable features of our model: First, it breaks the ergodicity and the VACF oscillates asymptotically, even though the nonergodic strength [\refeq{non-ergodic-strength}] vanishes.
This behavior is observed for any finite mass of the tagged particle and any finite coupling constant between the chain and the particle.
We point out that the failure of the nonergodic strength in the present model is due to the abuse of the final value theorem in \refeq{eqn:10}. The theorem holds only if $a(t)$ converges to a final value. Clearly, this is not the case for the present model. Finding a more robust criterion for the ergodicity remains as an interesting future problem.

Second, we found a localized mode with an isolated frequency from the continuous phonon spectrum, and that the existence of such a localized mode breaks the ergodicity of the system.
We should mention that some physical models have been proposed, which exhibit similar localized and isolated modes: In Ref.~\cite{Bao2017}, a finite-size system has been proposed, while in Ref.~\cite{Lee-Florencio-Hong1989} a harmonic oscillator chain system with a light impurity has been considered.
In both cases, however, the relation between the existence of such localized mode and the ergodicity breaking was not focused.

The memory function of the present model does not decay, but oscillates asymptotically. It should be pointed out that this behavior violates one of the three conditions proposed in Ref.~\cite{Plyukhin2011} as necessary conditions so that the GLE becomes {\rm physical}. The conditions are, (i)~the Laplace transform of the memory function should vanish in the large-$z$ limit, $\lim_{z \to \infty} \overline{\italic{\Gamma}}(z) = 0$, (ii)~the memory function should decay in the long time limit, $\lim_{t \to \infty} \itGamma(t) = 0$, and (iii)~the magnitude of the memory function should be smaller than the initial value, $|\itGamma(t)| < |\itGamma(0)|$. Although the memory function in the present study is derived analytically from a microscopic physical model and satisfies the conditions~(i) and (iii), it breaks the condition~(ii). This fact tells us that the condition~(ii) is {\rm not} necessary for the GLE to be physical.

\section*{ACKNOWLEDGEMENTS}
 The authors thank T.~Shirai for stimulating discussion and comments, and D.~Adachi and M.~Suzuki for careful reading of the manuscript.  F.I. is supported by the Japan Society for the Promotion of Science through the Program for Leading Graduate Schools (MERIT).

\appendix
\section{Total shift sampling}\label{sect:app}
\subsection{Algorithm}
We develop a method to perform sampling for a harmonic oscillator chain, called the ``total shift sampling.''
Our purpose is to generate configurations according to the Boltzmann distribution, Eq.~(\ref{eqn:boltzmann}).
Since the distribution function can be factorized into one-dimensional Gaussian distributions for $x(0)$, $\dot{x}(0)$, and $\{\dot{q}_n(0)\}$, generation of these random variables is straightforward.
The remaining problem is thus generating $\{q_n(0)\}$ according to the probability distribution~(\ref{eqn:boltzmann2}).

Here, we consider a finite chain of length $N$ with periodic boundary conditions, $q_{i+N} = q_i$. For simplicity, we drop `(0)' and renumber the variables as $q_1,q_2,\cdots,q_N$ and set $q_1=0$.
The joint probability distribution function is then written as
\begin{align}
  P_q(\{q_n\}_{n=2}^{N}) \propto e^{-\frac{1}{2\sigma^2} \sum_{n=1}^{N} (q_{n+1} - q_n)^2}
  \label{eqn:gaussian}
\end{align}
with $\sigma^2 = k_{\rm B}T$.

Sampling from the above distribution can be achieved according to the following procedure: First, we sample $N$ real random numbers, $\{r_n\}_{n=1}^N$, from the Gaussian distribution with variance $\sigma^2$. Then, we determine $\{q_n\}_{n=2}^N$ recursively as
\begin{align}
q_{n+1} & = q_{n} + r_{n} -  \frac{1}{N} \sum_{\ell = 1}^{N} r_\ell
= \sum_{j=1}^{n} r_j - \frac{n}{N} \sum_{\ell = 1}^{N} r_{\ell}.
\end{align}
Note that the periodic boundary condition
\begin{align}
q_{N+1} & = q_{N} + r_{N} -  \frac{1}{N} \sum_{\ell = 1}^{N} r_\ell
= \sum_{j=1}^{N} r_j - \frac{N}{N} \sum_{\ell = 1}^{N} r_{\ell} = 0
\end{align}
is satisfied by construction.

Then, $\{q_n\}_{n=2}^{N}$ satisfy the distribution~\refeq{eqn:gaussian}. It is easy to confirm the marginal distribution of a particular variable, but the joint probability distribution is not trivial.

\subsection{Validity of total shift sampling}
 We proof the validity of our algorithm. We can represent the distribution obtained by the above procedure as
 \begin{align}
   \begin{split}
& P_q(\{q_n\}_{n=2}^{N}) \propto \int \cdots \int  \prod_{n = 1}^{N} dr_n \\
& \quad \times \, \delta \bigl(q_{n+1} - ( q_{n} + r_{n} -  \frac{1}{N}  \sum_{\ell = 1}^{N} r_\ell ) \bigr) e^{-\frac{1}{2\sigma^2} \sum_{i=1}^{N} r^2_i},
  \end{split}
\end{align}
where $\delta(\cdot)$ is the Dirac delta function. We rewrite the Dirac delta function in the right-hand side as
\begin{align}
  \begin{split}
&\int \cdots \int  \prod_{n = 1}^{N} dr_n \, \delta \bigl(q_{n+1} - ( q_{n} + r_{n} -  \frac{1}{N}  \sum_{\ell = 1}^{N} r_\ell ) \bigr)  e^{-\frac{1}{2\sigma^2} \sum_{i=1}^{N} r^2_i} \\
& =  \int \cdots \int  \biggl\{\prod_{n = 1}^{N} dr_n \, d k_n \biggr\} \\
& \qquad \times e^{\sum_{n=1}^{N} \bigl[-ik_n\bigl(q_{n+1} - ( q_{n} + r_{n} -  \frac{1}{N}  \sum_{\ell = 1}^{N} r_\ell ) \bigr)- \frac{1}{2\sigma^2} r^2_n \bigr]} .
  \end{split}
\end{align}
The exponent can be transformed as
\begin{align}
  \begin{split}
& \sum_{n=1}^{N} \bigl[-ik_n\bigl(q_{n+1} - ( q_{n} + r_{n} -  \frac{1}{N}  \sum_{\ell = 1}^{N} r_\ell ) \bigr)- \frac{1}{2\sigma^2} r^2_n\bigr]\\
& = \sum_{n=1}^{N} \bigl[-ik_n\bigl(q_{n+1} - q_{n} \bigr)\bigr] \\ 
& \quad - \sum_{n=1}^{N} \frac{1}{2\sigma^2}\biggl[ r_n - \sigma^2 \bigl\{ ik_n - i\frac{1}{N}\sum_{n=1}^{N}k_n  \bigr\} \biggr]^2 \\
& \quad - \frac{\sigma^2}{2} \sum_{n=1}^{N}k^2_n + \frac{\sigma^2}{2N} \biggl(\sum_{n=1}^{N}k_n \biggr)^2.
    \end{split}
\end{align}
By performing the Gaussian integration with respect to $\{ r_n \}_{n=1}^{N}$, we obtain
\begin{align}
  \begin{split}
& P_q(\{q_n\}_{n=2}^{N}) \propto \int \cdots \int  \biggl\{\prod_{n = 1}^{N} dk_n \biggr\} \\
& \quad \times  e^{\sum_{n=1}^{N} \bigl[-ik_n   \bigl(q_{n+1} - q_{n} \bigr)\bigr] - \frac{\sigma^2}{2}\sum_{n=1}^{N} k^2_n + \frac{\sigma^2}{2N}\biggl(\sum_{n=1}^{N}k_n \biggr)^2 } .
  \end{split}
\end{align}
Then, we introduce a new variable, $Q = \sum_{n=1}^{N} k_n$, in the right-hand side as
\begin{align}
  \begin{split}
& \int \cdots \int dQ  \biggl\{\prod_{n = 1}^{N} d k_n \biggr\} \delta(Q - \sum_{n=1}^{N} k_n )\\
& \quad \times  e^{\sum_{n=1}^{N} \bigl[-ik_n\bigl(q_{n+1} - q_{n} \bigr)\bigr] - \frac{\sigma^2}{2}\sum_{n=1}^{N}k^2_n + \frac{\sigma^2}{2N}Q^2 } \\
= & \int \cdots \int  dQ \, d\eta \biggl\{\prod_{n = 1}^{N} d k_n \biggr\}  \\
\times & e^{-i\eta\bigl( Q - \sum_{n=1}^{N} k_n \bigr) + \sum_{n=1}^{N} \bigl[-ik_n\bigl(q_{n+1} - q_{n} \bigr)\bigr] - \frac{\sigma^2}{2}\sum_{n=1}^{N}k^2_n + \frac{\sigma^2}{2N}Q^2 } .
  \end{split}
\end{align}
We evaluate the exponent as
\begin{align}
  \begin{split}
& -i\eta\bigl( Q - \sum_{n=1}^{N} k_n \bigr) + \sum_{n=1}^{N} \bigl[-ik_n\bigl(q_{n+1} - q_{n} \bigr)\bigr]  \\
& \qquad - \frac{\sigma^2}{2}\sum_{n=1}^{N}k^2_n + \frac{\sigma^2}{2N}Q^2 \\
& =  -\frac{\sigma^2}{2}\sum_{n=1}^{N} \biggl[  k_n + i\frac{1}{\sigma^2}\bigl(q_{n+1} - q_{n} - \eta \bigr) \biggr]^2 \\
& \qquad - \frac{1}{2\sigma^2}\sum_{n=1}^{N}\bigl(q_{n+1} - q_{n} - \eta \bigr)^2  -i\eta Q + \frac{\sigma^2}{2N}Q^2 .
  \end{split}
\end{align}
After performing the Gaussian integration with respect to $\{k_n\}$, we have
\begin{align}
P_q(\{q_n\}_{n=2}^{N}) \propto \int \int  dQ \, d\eta \, e^{- \frac{1}{2\sigma^2}\sum_{n=1}^{N}\bigl(q_{n+1} - q_{n} - \eta \bigr)^2  -i\eta Q + \frac{\sigma^2}{2N}Q^2 }.
\end{align}
The exponent is further transformed as
\begin{align}
  \begin{split}
& - \frac{1}{2\sigma^2}\sum_{n=1}^{N}\bigl(q_{n+1} - q_{n} - \eta \bigr)^2  -i\eta Q + \frac{\sigma^2}{2N}Q^2 \\
& \quad =  - \frac{1}{2\sigma^2}\sum_{n=1}^{N}\bigl(q_{n+1} - q_{n}\bigr)^2 + \frac{\eta}{\sigma^2} \sum_{n=1}^{N}\bigl(q_{n+1} - q_{n}\bigr) \\
& \qquad  -\frac{N}{2\sigma^2}\eta^2  -i\eta Q + \frac{\sigma^2}{2N}Q^2 \\
& \quad =  - \frac{1}{2\sigma^2}\sum_{n=1}^{N}\bigl(q_{n+1} - q_{n}\bigr)^2 -\frac{N}{2\sigma^2}(\eta  +\frac{i\sigma^2}{N} Q)^2
  \end{split}
\end{align}
Here, we use the periodic boundary condition, $\sum_{n=1}^{N}(q_{n+1} - q_{n}) = q_{N+1} - q_1 = 0$.
Finally, we obtain the joint distribution as
\begin{align}
  \begin{split}
    P_q(\{q_n\}_{n=2}^{N})
& \propto \int \int  dQ \, d\eta \, e^{- \frac{1}{2\sigma^2}\sum_{n=1}^{N}\bigl(q_{n+1} - q_{n}\bigr)^2 - \frac{1}{2N\sigma^2}\bigl(\eta - \frac{i\sigma^2}{N}Q\bigr)^2} \\
& \propto e^{- \frac{1}{2\sigma^2}\sum_{n=1}^{N}\bigl(q_{n+1} - q_{n}\bigr)^2}.
  \end{split}
\end{align}

\bibliography{biblist}
\end{document}